\documentclass[prd,amsmath,amssymb,showpacs]{revtex4}
\usepackage{graphicx}
\usepackage{color}
\usepackage[colorlinks=true,linkcolor=red]{hyperref}
\usepackage{hyperref}

\begin{document}
\title{Phantom energy from graded algebras}
\author{Max Chaves}
\email{mchaves@cariari.ucr.ac.cr} \affiliation{Escuela de Fisica
\\ Universidad de Costa Rica, San Jose, Costa Rica}
\author{Douglas Singleton}
\email{dougs@csufresno.edu} \affiliation{Physics Dept., CSU Fresno
Fresno, CA 93740-8031 \\ and \\
Universidad de Costa Rica, San Jose, Costa Rica}

\date{\today}

\begin{abstract}
We construct a model of phantom energy using the graded Lie
algebra SU(2/1). The negative kinetic energy of the phantom field emerges
naturally from the graded Lie algebra, resulting in an equation of state with $w<-1$. The model
also contains ordinary scalar fields and anti-commuting (Grassmann) vector 
fields which can be taken as two component dark matter. A potential term is generated 
for both the phantom fields and the ordinary scalar fields via a postulated 
condensate of the Grassmann vector fields. Since the phantom energy and dark matter
arise from the same Lagrangian the phantom energy and dark matter of this model 
are coupled via the Grassman vector fields. In the model presented here 
phantom energy and dark matter come from a gauge principle rather than being introduced
in an {\it ad hoc} manner.      
\end{abstract}


\maketitle


\section{Introduction}

Graded Lie algebras or Lie superalgebras ({\it i.e.} algebras having commuting
and anti-commuting generators) were at one time considered as models for a
more complete unified electroweak theory \cite{dondi} \cite{neeman} \cite{fairlie}
\cite{squires} \cite{taylor1} as well as Grand Unified Theories 
\cite{taylor}. Such graded algebras had many attractive features such as including
both vector and scalar bosons within the same theory, fixing the Weinberg angle,
and in some formulations the mass of the Higgs. However these graded algebras had
shortcomings \cite{eccle} \cite{eccle1} such as giving rise to negative kinetic energy terms
for some of the gauge fields when the graded trace or supertrace was used. 

In this paper we point out that this negative kinetic energy feature 
of the original graded algebras can be used to 
construct a model for phantom energy \cite{caldwell}. In addition to the phantom field
there are other fields which arise in this model which could act as dark matter. The
advantage of the combined phantom energy/dark matter model presented here is that 
it is derived from a modified gauge principle ({\it i.e.} the gauge principle applied
to graded algebras) rather than being introduced phenomenologically. This feature
fixes the parameters, such as the coupling between the phantom energy and dark
matter, that are free in more phenomenological models.  

Phantom energy is a form of dark energy which has a ratio of density to pressure less than
$-1$ {\it i.e.} $w = p / \rho < -1$.   Dark energy in general is a cosmological
``fluid'' with $w < -1/3$, which gives rise to an accelerated 
cosmological expansion. Dark energy was proposed to explain the accelerated
expansion that was observed in studies of distance type Ia supernova \cite{riess} \cite{riess1}
\cite{perlmutter}. There are various proposals as to the nature of dark energy
which include a small, positive cosmological constant, quintessence models \cite{zlatev},
brane world models \cite{ddg} \cite{deff}, Chaplyin gas \cite{kam}, k-essense \cite{gonzalez1}, 
axionic tensor fields \cite{gonzalez2} and others. A good review of the
the current hypotheses as to the nature of dark energy and a guide to the literature can be 
found in reference \cite{sahni}. Phantom energy is simply an extreme form of dark energy.  The simplest 
model for phantom energy involves a scalar field with a negative kinetic energy term \cite{caldwell} 
\begin{equation}
\label{scalar-pe}
{\mathcal L}_p = -\frac{1}{2} (\partial _\mu \phi) (\partial ^\mu \phi) - V(\phi)
\end{equation}
The negative sign in front of the kinetic energy term makes this an
unusual field theory. Such theories with negative kinetic energies have been
investigated theoretically starting with \cite{bronn} \cite{ellis} where they were used to
study wormhole solutions. Other papers considering scalar fields with
negative kinetic energies can be found in \cite{nega-ke} \cite{picon}
\cite{lobo} \cite{sushkov}. However these theoretical
studies did not garner much attention because of various problems with
negative kinetic energies. Quantum mechanically such a field theories 
violate either conservation of probability or it has no
stable vacuum state due to an energy density that is unboundedly negative.
Nevertheless such unusual field theories are not absolutely ruled out
\cite{caldwell}, but one can place significant constraints on them \cite{cline}. 
Despite the theoretical problems of a scalar field with a negative kinetic energy
term the reason to consider such a strange field theory is that recent
observations give  $-1.48 < w < -0.72$ \cite{hann} \cite{melch} \cite{knop} \cite{sper}
\cite{tegmark} and thus favor $w < -1$. A very recent
comparison of data from various sources can be found in \cite{jass} 

The result $w<-1$ coming from the Lagrangian in \eqref{scalar-pe} depends not only on the 
negative kinetic energy term, but also requires that the potential, $V(\phi )$, be present
and satisfy some conditions. From \eqref{scalar-pe} one can calculate $p$ and $\rho$
in the standard way as
\begin{equation}
\label{p-rho}
\rho = T_{00} = -\frac{1}{2}{\dot \phi} ^2 + V(\phi ) \; ; \hspace{2cm}
p = - T_{ii} = -\frac{1}{2}{\dot \phi} ^2 - V(\phi )      
\end{equation}
where it is assumed that the scalar field is spatially homogeneous enough so that only the 
time variation is important. Using \eqref{p-rho} to calculate $w$ gives
\begin{equation}
\label{w}
w = \frac{p}{\rho} = \frac{-\frac{1}{2}{\dot \phi} ^2 - V(\phi )}{-\frac{1}{2}{\dot \phi} ^2 + V(\phi )}
\end{equation}
If $V (\phi ) >0$ and satisfies $ \sqrt { 2 V (\phi ) } > | {\dot \phi} |$ then one has $w <-1$. 
We will show that it is possible, using graded algebras, to construct a field theory that satisfies
these conditions and so gives rises to phantom energy with $w <-1$. Unlike other models, the
negative kinetic term comes from the structure of the graded algebras rather than being put
in by hand. In addition there are other fields which could play the role of dark matter. 

\section{SU(2/1) algebra}

We briefly review the graded algebra SU(2/1), which we will use to construct the phantom energy
model in the next section. The basic idea of using graded algebras to give phantom energy
works for larger graded algebras like SU(N/1) with $N > 2$. We have taken SU(2/1) for
simplicity.

We use the representation for SU(2/1) which consists of the following eight 
$3 \times 3$ 
\begin{align*}
EVEN: \quad T_{1}  &  =\frac{1}{2}\left(
\begin{array}
[c]{ccc}%
0 & 1 & 0\\
1 & 0 & 0\\
0 & 0 & 0
\end{array}
\right)  ,\quad T_{2}=\frac{1}{2}\left(
\begin{array}
[c]{ccc}%
0 & -i & 0\\
i & 0 & 0\\
0 & 0 & 0
\end{array}
\right)  ,\quad T_{3}=\frac{1}{2}\left(
\begin{array}
[c]{ccc}%
1 & 0 & 0\\
0 & -1 & 0\\
0 & 0 & 0
\end{array}
\right) \quad T_{8}=\frac{1}{2}\left(
\begin{array}
[c]{ccc}%
1 & 0 & 0\\
0 & 1 & 0\\
0 & 0 & 2
\end{array}
\right)
\\
ODD: \quad T_{4}  &  =\frac{1}{2}\left(
\begin{array}
[c]{ccc}%
0 & 0 & 1\\
0 & 0 & 0\\
1 & 0 & 0
\end{array}
\right)  ,\quad T_{5}=\frac{1}{2}\left(
\begin{array}
[c]{ccc}%
0 & 0 & -i\\
0 & 0 & 0\\
i & 0 & 0
\end{array}
\right)  ,\quad T_{6}=\frac{1}{2}\left(
\begin{array}
[c]{ccc}%
0 & 0 & 0\\
0 & 0 & 1\\
0 & 1 & 0
\end{array}
\right) , \quad T_{7} =\frac{1}{2}\left(
\begin{array}
[c]{ccc}%
0 & 0 & 0\\
0 & 0 & -i\\
0 & i & 0
\end{array}
\right)  .
\end{align*} 
Except for $T_8$ this is the standard, fundamental 
representation of SU(3). The matrices on the first line above ({\it i.e.} 
$T_1, T_2, T_3, T_8$) are the even generators, and those on the second line
({\it i.e.} $T_4, T_5, T_6, T_7$) are odd generators. The even generators satisfy 
commutation relationships among themselves which can be written symbolically as
$[EVEN, EVEN] = EVEN$. Mixtures of even and odd generators satisfy commutators of
the form $[EVEN, ODD] = ODD$. Finally the odd generators satisfy anti-commutation
relationships of the form $\{ ODD, ODD \} = EVEN$. The further details of the SU(2/1)
graded algebra can be found in the paper by Dondi and Jarvis \cite{dondi} or
in Ecclestone \cite{eccle} \cite{eccle1}. The odd generators above are different than those
usually taken in the literature. The connection of the odd generators above with
those in \cite{dondi} is given by ${\bar Q} ^1 , Q_1 = T_4 \pm i T_5$
and ${\bar Q} ^2 , Q_2 = T_6 \pm i T_7$. In the rest of the article we will use the
convention that generators with indices from the middle of the alphabet ($i, j, k$)
are the even generators, $T_1, T_2, T_3, T_8$, while indices from the beginning of the 
alphabet ($a, b, c$) are the odd generators $T_4, T_5, T_6, T_7$.

For the graded algebra one replaces the concept of the trace by the supertrace. 
For SU(2/1) this means that one writes some general element of the group as
\begin{align*}
\quad M =\left(
\begin{array}
[c]{cc}%
A_{2 \times 2} & B_{2 \times 1} \\
C_{1 \times 2} & d_{1 \times 1} \\
\end{array}
\right)  .
\end{align*} 
The subscripts indicate the size of the sub-matrix. The supertrace is now defined as
\begin{equation}
\label{supertrace}
{\operatorname {str}}(M) = {\operatorname {tr}} [A] - {\operatorname {tr}}[d]
\end{equation}
which differs from the regular trace due to the minus sign in front of $d$.

In the next section we will need the supertraces of the various products of the
eight generators $(T_i , T_a)$, thus we collect these results here. For products of
even generators we have
\begin{equation}
\label{even}
{\operatorname {str}}(T_i T_j) = \delta _{ij} \frac{1}{2} \quad \text{except} \quad 
{\operatorname {str}}(T_8 T_8) = -\frac{1}{2}
\end{equation}
for the odd generators we have
\begin{equation}
\label{odd}
{\operatorname {str}} (T_4 T_5) = -{\operatorname {str}}(T_5 T_4) = \frac{i}{2} , \quad 
{\operatorname {str}} (T_6 T_7) = -{\operatorname {str}}(T_7 T_6) = \frac{i}{2}
\end{equation}
All other supertraces of the product of two matrices are zero.

\section{Phantom energy from SU(2/1)}

In \cite {dondi} vector fields were associated with the even
generators and scalar fields with the odd generators as
\begin{equation}
\label{graded-alg}
A_\mu = i g A_\mu ^i T^{even} _i \; \; , \quad \phi  = -g \varphi ^a  T^{odd} _a
\end{equation}
The fields $A_\mu ^i$ are regular commuting fields while $\varphi ^a$ are Grassmann fields.
In block form one can write \eqref{graded-alg} as
\begin{align*}
\quad A_M =\left(
\begin{array}
[c]{ccc}%
A_\mu ^3 + A_\mu ^8 & A_\mu ^1 - i A _\mu ^2 & \varphi ^4 - i \varphi ^5\\
A_\mu ^1 + i A_\mu ^2 & -A_\mu ^3 + A_\mu ^8 & \varphi ^6 - i \varphi^7\\
\varphi ^4 + i \varphi ^5 & \varphi ^6 + i \varphi ^7 & 2 A_\mu ^8
\end{array}
\right)  .
\end{align*} 
In this fashion, {\it and by using the regular trace},  Dondi and 
Jarvis \cite{dondi} showed that the Lagrangian 
\begin{equation}
\label{g-lag}
{\mathcal L} = \frac{1}{2 g^2} {\operatorname {tr}} ( F_{MN} F^{MN} ) , \quad F_{MN} = \partial_M A_N  
-\partial _N A_M + [A_M, A_N] ,
\end{equation}
reduced to an $SU(2) \times U(1)$ Yang-Mills Lagrangian for $A_\mu$ and a 
Higgs-like Lagrangian for $\phi$. In \eqref{g-lag} we use a different overall
sign for the Lagrangian as compared to \cite{dondi}. This comes because we 
have chosen different factors of $i$ in the vector potentials defined 
below in \eqref{graded-alg2}. This was a more unified electroweak
theory since $SU(2/1)$ is simple so that the Weinberg angle was fixed rather
than being a parameter. However, if in \eqref{g-lag} one used the correct
SU(2/1) invariant supertrace then the Yang-Mills part of the reduced Lagrangian 
would have the wrong sign for the kinetic term for the $U(1)$ gauge field
and the kinetic energy term for the scalar field
would be lost. In \cite{eccle} \cite {eccle1} this was used to develop arguments against the 
use of SU(2/1) as a unified electroweak theory. 

Here we would like to use these apparent negative features of the graded algebras
to construct a model for phantom energy. Instead of making the association between 
even/odd generators and vector/scalar fields made in \eqref{graded-alg} we will make
the opposite choice
\begin{equation}
\label{graded-alg2}
A_\mu = i g A_\mu ^a T^{odd} _a \; \; , \quad \phi = - g \varphi  ^i T^{even} _i
\end{equation}  
Because of the reversal of roles relative to \eqref{graded-alg} the fields $A_\mu ^a$ 
are Grassmann fields while $\varphi ^i$ are regular, commuting fields. Then taking the
correct SU(2/1) invariant supertrace we find that one of the scalar fields develops
a negative kinetic energy term in addition to having a potential term which is positive
definite. Thus the graded algebra gives rise to a phantom field. 

With the choice in \eqref{graded-alg2} the Lagrangian in \eqref{g-lag} reduces as
follows 
\begin{equation}
\label{g-lag2}
{\mathcal L} = \frac{1}{2 g^2} {\operatorname{str}} ( F_{MN} F^{MN} ) = 
\frac{1}{2g^{2}}{\operatorname{str}}\left[  \left(  \partial_{\lbrack\mu}A_{\nu]}+[A_{\mu},A_{\nu}]\right)
^{2}\right] +\frac{1}{g^{2}}{\operatorname{str}}\left[  \left(  \partial
_{\mu}\phi  + [A_{\mu},\phi ]\right)  ^{2}\right]   
\end{equation}
We have introduced the notation 
$\partial_{\lbrack\mu}A_{\nu]} = \partial _\mu A_\nu - \partial _\nu A_\mu$.  Note
that in comparison to other works such as \cite{dondi} and \cite{eccle} \cite{eccle1} we have not
introduced extra Grassmann coordinates, $\zeta^\alpha$ in addition to the normal
Minkowski coordinates $x^\mu$. Thus in \cite{dondi} and \cite{eccle} \cite{eccle1} coordinates and 
indices ran over six values -- four Minkowski and two Grassmann. The final result in \eqref{g-lag2} 
can be obtained from \cite{dondi} by dropping the Grassmann coordinates. 

We first focus  on the scalar term in \eqref{g-lag2}. Inserting $\phi$ and $A_\mu$ from 
\eqref{graded-alg2} into the last term in \eqref{g-lag2} we find
\begin{eqnarray}
\label{scalar}
{\mathcal L}_S & = & \frac{1}{g^{2}}{\operatorname{str}}\left[
\left(  \partial_{\mu}\phi+[A_{\mu},\phi]\right)  ^{2}\right] \nonumber \\
& = & {\operatorname{str}}\left[  \left(  \partial_{\mu}\varphi
^{8}T_{8}+igA_{\mu}^{a}\varphi^{8}[T_{a},T_{8}]\right)  ^{2}\right]
+ {\operatorname{str}}\left[  \left(  \partial_{\mu}\varphi
^i T_i +igA_{\mu}^{a}\varphi^i[T_{a},T_i]\right)  ^{2}\right]
\end{eqnarray} 
We now show that the first term in \eqref{scalar} takes the form of a phantom energy field. 
Expanding the first term in \eqref{scalar} gives
\begin{eqnarray}
\label{phantom}
\mathcal{L}_{Ph}  & = & {\operatorname{str}}\left[  \left(
\partial_{\mu}\varphi^{8}T_{8}+igA_{\mu}^{4}\varphi^{8}[T_{4},T_{8}]+igA_{\mu
}^{5}\varphi^{8}[T_{5},T_{8}] + igA_{\mu}^{6}\varphi^{8}[T_{6},T_{8}]
+ igA_{\mu}^{7}\varphi^{8}[T_{7},T_{8}] \right)  ^{2}\right] \nonumber \\
& = & {\operatorname{str}}\left[  \left(  \partial_{\mu}\varphi
^{8}T_{8}-gA_{\mu}^{4}\varphi^{8}T_{5}/2+gA_{\mu}^{5}\varphi^{8}%
T_{4}/2 -gA_{\mu}^{6}\varphi^{8}T_{7}/2 + gA_{\mu}^{7}\varphi^{8}%
T_{6}/2\right)  ^{2}\right]
\end{eqnarray}
We have used the representation of the SU(2/1) matrices from the previous 
section to evaluate the commutators. Using the supertrace results from \eqref{even}
and \eqref{odd} the expression in \eqref{phantom} yields
\begin{eqnarray}
\label{phantom2}
\mathcal{L}_{Ph}  &  = &-\frac{1}{2}(\partial_{\mu}\varphi^{8})^{2}+\frac{1}%
{4}g^{2}(\varphi^{8})^{2}{\operatorname{str}} \left[ \left( A_{\mu}
^{4}T_{5}-A_{\mu}^{5}T_{4} + A_{\mu}^{6}T_{7}-A_{\mu}^{7}T_{6} \right) ^2 \right] \nonumber \\
& = & -\frac{1}{2}(\partial_{\mu}\varphi^{8})^{2}- \frac{i}{8}g^{2}(\varphi
^{8})^{2} \left( A_{\mu}^{5}A^{4 \mu} - A_{\mu}^{4}A^{5 \mu} + A_{\mu}^{7} A^{6 \mu} 
- A_{\mu}^{6}A^{7 \mu} \right) \\
&  = & -\frac{1}{2}(\partial_{\mu}\varphi^{8})^{2}- \frac{1}{16}g^{2}(\varphi
^{8})^{2}\left( A_{\mu}^{+}A^{- \mu} - A_{\mu}^{-}A^{+ \mu} + B_{\mu}^{+}B^{- \mu}
- B_{\mu}^{-}B^{+ \mu} \right) \nonumber 
\end{eqnarray}
with $A_{\mu}^\pm=A_{\mu}^{4} \pm iA_{\mu}^{5}$
and $B_{\mu}^\pm =A_{\mu}^{6} \pm iA_{\mu}^{7}$. Both $A_{\mu} ^\pm$
and $B_{\mu} ^\pm$ are Grassmann so the last line in \eqref{phantom2}
can be written
\begin{equation}
\label{phantom3}
\mathcal{L}_{Ph} = -\frac{1}{2}(\partial_{\mu}\varphi^{8})^{2}- \frac{1}{8}g^{2}(\varphi
^{8})^{2}\left( A_{\mu}^{+}A^{- \mu} + B_{\mu}^{+}B^{- \mu} \right)
\end{equation}
This is of the form of the phantom energy Lagrangian in \eqref{scalar-pe} but with the
potential involving not only the scalar field, $\varphi ^8$, but Grassmann vector fields,
$A_\mu ^\pm$ and $B_\mu ^\pm$. We will discuss these shortly. The minus sign in front of the kinetic 
energy term comes from taking the SU(2/1) invariant supertrace rather
than the ordinary trace (see the second supertrace result in \eqref{even}). 

We next focus on the other scalar fields, $\varphi ^i$, $i =1,2,3$ which come from the
second term in \eqref{scalar}. The calculation proceeds as in equations \eqref{phantom} - 
\eqref{phantom2} but with $\varphi ^8$ replaced by $\varphi ^i$, $i= 1, 2, 3$. For example
for $\varphi ^1$ \eqref{phantom} becomes
\begin{equation}
\label{dm}
\mathcal{L}_{\varphi ^1}  = {\operatorname{str}}\left[  \left(  \partial_{\mu}\varphi
^{1}T_{1}+gA_{\mu}^{4}\varphi^{1}T_{7}/2-gA_{\mu}^{5}\varphi^{1}%
T_{6}/2 -gA_{\mu}^{6}\varphi^{1}T_{5}/2 + gA_{\mu}^{7}\varphi^{1}%
T_{4}/2\right)  ^{2}\right] 
\end{equation}
and \eqref{phantom2} becomes
\begin{eqnarray}
\label{dm2}
\mathcal{L}_{\varphi^1} & = & \frac{1}{2}(\partial_{\mu}\varphi^{1})^{2}- \frac{1}{16}g^{2}(\varphi
^{1})^{2}\left( A_{\mu}^{+}A^{- \mu} - A_{\mu}^{-}A^{+ \mu} + B_{\mu}^{+}B^{- \mu}
- B_{\mu}^{-}B^{+ \mu} \right) \nonumber \\
 & = & \frac{1}{2}(\partial_{\mu}\varphi^{1})^{2}- \frac{1}{8}g^{2}(\varphi
^{1})^{2}\left( A_{\mu}^{+}A^{- \mu} + B_{\mu}^{+}B^{- \mu} \right)
\end{eqnarray} 
There are two keys points: the kinetic term for $\varphi ^1$ is positive since 
${\operatorname{str}} (T_1 T_1) = + 1/2$, and the potential term is the same as
for $\varphi ^8$. The other two even scalar fields follow
the same pattern so that in total one can write
\begin{equation}
\label{dm3}
\mathcal{L}_{DM} = \frac{1}{2}(\partial_{\mu}\varphi^{i})^{2}- \frac{1}{8}g^{2}(\varphi
^{i})^{2}\left( A_{\mu}^{+}A^{- \mu} + B_{\mu}^{+}B^{- \mu} \right)
\end{equation}
where $i$ is summed from 1 to 3. Thus the total scalar field Lagrangian resulting from
\eqref{scalar} is the sum of \eqref{phantom3} and \eqref{dm3}. The scalar field in 
\eqref{phantom3} has the ``wrong'' sign for the kinetic term and acts as a phantom
field. The scalar fields in \eqref{dm3} are ordinary scalar field which we will interpret 
as a dark matter candidate. The phantom field and dark matter fields are coupled
through the $A^\pm _\mu$ and $B^\pm _\mu$ fields. Thus our model provides a coupling 
between phantom energy and dark matter. Other models have been considered \cite{cai} where
there is coupling between dark/phantom energy and dark matter.

We will now examine the Grassmann vector fields, $A^4 _\mu , 
A^5 _\mu, A^6 _\mu , A^7 _\mu$. The final Lagrangian for these fields will
have a nonlinear interaction between the $A^{\pm}_\mu$ and $B^{\pm}_\mu$ fields.
In analogy with QCD we argue that these fields form permanently confined condensates like
$\langle A^4 _\mu A^5 _\mu \rangle$ or $\langle A^+ _\mu A^- _\mu \rangle$. 
These then supply potential (mass-like)
terms for the phantom energy and scalar fields of \eqref{phantom3} and \eqref{dm3}.
This also avoids violation of the spin-statistics theorem since these condensates
have bosonic statistics (they are composed of two Grassmann fields) and integer 
spin (they are composed of two integer spin fields). 
Having a potential term is crucial for the interpretation of $\varphi ^8$ as a
phantom energy field, since for a massless, non-interacting scalar field reversing the
sign of the kinetic energy term does not lead a phantom field with $w<-1$ as can be seen
from \eqref{w} if $V( \phi) =0$.
From \eqref{g-lag2} the vector part of the Lagrangian can be expanded as
\begin{eqnarray}
\label{vector}
\mathcal{L}_V & = & \frac{1}{2g^{2}}{\operatorname{str}}
\left[  \left(  \partial_{\lbrack\mu}A_{\nu]}+[A_{\mu},A_{\nu}]\right)^{2}\right]
=  - \frac{1}{2}{\operatorname{str}}
\left[  \left(  \partial_{\lbrack\mu}A^a _{\nu]} T_a + ig A^a _\mu A^b _\nu \{ T_a,T_b \} \right)
^{2}\right] \nonumber \\
& = &  - \frac{1}{2}{\operatorname{str}}
\left[  \left(  \partial_{\lbrack\mu}A^a _{\nu]} T_a \right) ^2 \right] 
+ \frac{g^2}{2} {\operatorname{str}}\left[ \left( A^a _\mu A^b _\nu \{ T_a,T_b \} \right)
^{2}\right] = \mathcal{L}_{V1} + \mathcal{L}_{V2}
\end{eqnarray}
The commutator has become an anticommutator due to the Grassmann nature of the $A^a _\mu$'s. Also
note that there is no cubic cross term between the derivative and anticommutator part. This
comes about since the anticommutator, $\{ T_a , T_b \}$ results in even generators, and 
the supertrace between odd and even generators vanishes. $\mathcal{L}_{V1}$ is
a kinetic term for the fields and $\mathcal{L}_{V2}$ a potential term. We will now consider each
of these in turn.  

The kinetic part can be written explicitly as
\begin{equation}
\label{v-kinetic}
\mathcal{L}_{V1} =  - \frac{1}{2}{\operatorname{str}}
\left[  \left(  \partial_{\lbrack\mu}A^4 _{\nu]} T_4 + \partial_{\lbrack\mu}A^5 _{\nu]} T_5
+\partial_{\lbrack\mu}A^6 _{\nu]} T_6 + \partial_{\lbrack\mu}A^7 _{\nu]} T_7\right) ^2 \right] 
\end{equation}
Due to the property of the supertrace of the odd generators given in \eqref{odd} it is
only the cross terms between $T_4, T_5$ and $T_6, T_7$ which survive. 
\begin{eqnarray}
\label{v-kinetic2}
\mathcal{L}_{V1} = - \frac{i}{2}
\left(  \partial_{\lbrack\mu}A^4 _{\nu]} \partial_{\lbrack\mu}A^5 _{\nu]} 
+\partial_{\lbrack\mu}A^6 _{\nu]} \partial_{\lbrack\mu}A^7 _{\nu]} \right)
= - \frac{1}{4}
\left(  \partial_{\lbrack\mu}A^- _{\nu]} \partial_{\lbrack\mu}A^+ _{\nu]} 
+\partial_{\lbrack\mu}B^- _{\nu]} \partial_{\lbrack\mu}B^+ _{\nu]} \right)
\end{eqnarray}
where we have used the anticommutating properties of the $A ^a _\mu$'s.
In the last step we have replaced the $A ^a _\mu$ by $A^\pm _\mu$ and
$B^\pm _\mu$. This kinetic part is reminiscent of the kinetic terms for
a charged ({\it i.e.} complex) vector field.

Next we work out the form of the interaction terms coming from 
${\mathcal L}_{V2}$. We do this explicitly for $A^4 _\mu$; the
results for the other vectors fields can be obtained in a similar manner.
The $A^a _\mu = A^4 _\mu$ part of $\mathcal{L}_{V2}$ expands like
\begin{equation}
\label{vector-int}
{\mathcal L}_{V2} = \frac{g^2}{2} {\operatorname{str}}\left[ \left( A^4 _\mu A^4 _\nu \{ T_4,T_4 \} 
+ A^4 _\mu A^5 _\nu \{ T_4,T_5 \} + A^4 _\mu A^6 _\nu \{ T_4,T_6 \}
+ A^4 _\mu A^7 _\nu \{ T_4,T_7 \}
\right)^{2}\right]
\end{equation}   
Using the explicit representations of the odd matrices we have $\{T_4 , T_4 \} =
(T_3 + T_8)/2$, $\{ T_4 , T_5 \} = 0$, $\{ T_4 , T_6 \} = T_1 /2 $, 
$\{ T_4 , T_7 \} = -T_2 / 2$. Squaring and using the supertrace results
of \eqref{even} one finds that \eqref{vector-int} becomes
\begin{equation}
\label{vector-int2}
{\mathcal L}_{V2} = \frac{g^2}{16}\left( A^4 _\mu A^6 _\nu A^{4 \mu} A^{6 \nu} 
+ A^4 _\mu A^7 _\nu A^{4 \mu} A^{7 \nu} \right)
\end{equation}
Note that there is no quartic term in $A^4 _\mu$ since the contributions from
$T_3$ and $T_8$ cancel. The contribution from $A^5 _\mu$ looks the same as
\eqref{vector-int2} but with $A^4 _\mu \rightarrow A^5 _\mu$. The $A^6 _\mu$
and $A^7 _\mu$ terms can be obtained by making the exchange $A^4 _\mu
\leftrightarrow A^6 _\mu$ and $A^5 _\mu \leftrightarrow A^7 _\mu$. Using the
Grassmann character of the $A^a _\mu$ 's one can see that the $A^4 _\mu$ and $A^6 _\mu$
contributions, and also the $A^5 _\mu$ and $A^7 _\mu$ contributions are the same. 
In total the interaction part of the vector Lagrangian can be written as
\begin{eqnarray}
\label{vector-int3}
{\mathcal L}_{V2} & = & \frac{g^2}{8}\left( A^4 _\mu A^6 _\nu A^{4 \mu} A^{6 \nu} 
+ A^4 _\mu A^7 _\nu A^{4 \mu} A^{7 \nu}  + A^5 _\mu A^6 _\nu A^{5 \mu} A^{6 \nu} 
+ A^5 _\mu A^7 _\nu A^{5 \mu} A^{7 \nu}\right) \nonumber \\
& = & \frac{g^2}{16} \left( A^+ _\mu B^+ _\nu A^{- \mu} B^{- \nu} + A^+ _\mu B^- _\mu
A^{- \nu} B^{+ \nu} \right)
\end{eqnarray}
In the last line we have written the interaction in terms of $A^{\pm} _\mu$ , $B^{\pm} _\mu$. 

The total Lagrangian for the vector Grassmann fields is, ${\cal L}_{V1} + {\cal L}_{V2}$, where
${\cal L}_{V1}$ is a kinetic term and  and ${\cal L}_{V2}$ gives a nonlinear
interaction term between $A^\pm _\mu$ and $B^\pm _\mu$. We assume that the interaction
is strong enough that the fields, $A^\pm _\mu$ and $B^\pm _\mu$ are permanently confined into 
condensates 
\begin{equation}
\label{vev}
\langle A^+ _\mu A^{- \mu} \rangle = \langle B^+ _\mu B^{- \mu} \rangle = v
\end{equation}
Given the symmetry between the $A^{\pm}_\mu$ and $B^{\pm} _\mu$ fields we have
taken them to have the same vacuum expectation value. This conjectured condensation 
is similar to the {\it gauge variant}, mass dimension 2 condensate, in regular
Yang-Mills theory, $\langle {\cal A}^a _\mu {\cal A}^{a \mu} \rangle$. Despite being
{\it gauge variant} this quantity has been shown \cite{boucaud} \cite{boucaud1} \cite{gubarev} \cite{gubarev1} to 
have real physical consequences in 
QCD. Here ${\cal A}^a _\mu$ is a normal SU(N) Yang-Mills field. In \cite{kondo} a BRST-invariant 
mass dimension 2 condensate was constructed which was a combination of the quadratic
gauge field term -- $\langle {\cal A}^a _\mu {\cal A}^{a \mu} \rangle$ --
plus a quadratic Fadeev-Popov \cite{fp} ghost field term -- $i \alpha \langle {\cal C}^a  {\bar{\cal C}}^a \rangle$ --
where $\alpha$ was a gauge parameter. In the Landau gauge, $\alpha =0$, this reduced to a
pure quadratic gauge field condensate $\langle {\cal A}^a _\mu {\cal A}^{a \mu} \rangle$.
Note that the ghost fields, ${\cal C}^a , {\bar{\cal C}}^a$, are bosonic (i.e. scalar) 
Grassman fields. This mass dimension 2 condensate gives the gluon a mass 
\cite{ds} \cite{ds1} \cite{shakin}. Estimates have been made for 
$\sqrt{\langle {\cal A}^a _\mu {\cal A}^{a \mu} \rangle}$ using lattice methods
\cite{boucaud} \cite{boucaud1} \cite{shakin}, analytical techniques \cite{dudal} or some mixture.
All these various methods give a condensate value in the range
$\sqrt{\langle {\cal A}^a _\mu {\cal A}^{a \mu} \rangle} \approx 1$ GeV.  
Given the similarities between the regular gauge field condensate 
of \cite{boucaud} \cite{boucaud1} \cite{gubarev} \cite{gubarev1} \cite{kondo} and that on the left
hand side of \eqref{vev} we can estimate the vacuum expectation value as
$v \approx 1$ GeV $^2$.  

Now inserting these vacuum expectation values from \eqref{vev}
into \eqref{phantom3} yields
\begin{equation}
\label{phantom4}
\mathcal{L}_{Ph} = -\frac{1}{2}(\partial_{\mu}\varphi^{8})^{2}- \frac{v}{4}g^{2}(\varphi
^{8})^{2}
\end{equation} 
This is of the form \eqref{scalar-pe} with $V(\varphi ^8) =  \frac{v}{4}g^{2}(\varphi
^{8})^{2}$. This will give phantom energy with $w < -1$ if
$\frac{g}{2} |\varphi ^8 | {\sqrt{2 v}} > | \dot \varphi ^8 |$.
If the vacuum expectation value, $v$, change over time it is possible
to cross into (out of) the phantom regime if $v$ increases (decreases).
Thus whether one has phantom energy or not would depend on the dynamical 
evolution of $v$. Such models, where one crosses the ``phantom divide'', have been
considered in \cite{feng1} \cite{phantom-divide} \cite{andri} \cite{noji}. Usually in such models it is the sign in front of the
kinetic energy term that is modified, whereas in the present case it is a modification
of the potential which causes the transition between phantom and non-phantom phases.
Further extensions of these ``quintom" models can be found in \cite{feng2} \cite{feng3} 
\cite{feng4} \cite{zhao} \cite{xia}. 

Inserting the vacuum expectation values into the Lagrangian for the scalar fields
$\varphi _1 , \varphi _2 , \varphi _3$, equation \eqref{dm3} becomes
\begin{equation}
\label{dm4}
\mathcal{L}_{DM} = \frac{1}{2}(\partial_{\mu}\varphi^{i})^{2}- \frac{v}{4}g^{2}(\varphi
^{i})^{2}
 \end{equation}
The Lagrangian for these fields is for a standard, non-interacting scalar with mass
$m= \frac{g}{2}{\sqrt{2 v}}$. These massive scalar fields could be taken as a
candidate for cold dark matter if $m$ (i.e. $v$) is chosen appropriately. For example,
using the similarity between the condensate of \eqref{vev} and the mass dimension
condensate of \cite{boucaud} \cite{boucaud1} \cite{gubarev} 
\cite{gubarev1} \cite{kondo} one might set $v \approx 1$ GeV $^2$.
This would given $m \approx 1$ GeV making $\varphi ^a$ a cold dark matter candidate.  

The original Lagrangian we started with in \eqref{g-lag2} has no coupling to the
usual Standard Model fields except through gravity. This would certainly explain
why these phantom energy and dark matter fields have not been seen since they
could only be detected through their gravitational influence. However if this
is the path nature chooses it would be hard if not impossible to get any kind
of experimental signal of these phantom energy/dark matter candidates. One could introduce
some effective coupling between the phantom energy/dark matter fields of \eqref{g-lag2}
and the usual Standard Model fields. More rigorously one might try to use some larger
SU(N/1) group, but having some of the vector fields be associated with the even generators
and some associated with the odd generators and similarly for the scalar fields. In this way
it might be possible to have a new kind of ``Grand Unified Theory": from a single Lagrangian 
one could have Standard Model gauge fields as well as new fields that would be phantom energy and 
dark matter candidates, instead of extra Grand Unified gauge bosons.

The Grassmann vector fields are an odd feature of this model since they would violate the
spin-statistics theorem. These Grassmann vector fields are similar to the Fadeev-Popov ghosts \cite{fp}
which are scalar fields with Fermi-Dirac statistics. The Fadeev-Popov ghosts however are
not real particles in that they never appear as asymptotic states. In order to avoid having
the Grassmann vector fields violate the spin-statistics theorem, we have postulated that
the composite states, $A^+ _\mu A^{- \mu}$ and $B^+ _\mu B^{- \mu}$ are permanently confined so that
the particles associated with $A^\pm _\mu$ and $B^\pm _\mu$ never appear as asymptotic states.
Since the composites are ordinary fields (integer spin with bosonic statistics) violation
of the spin-statistics theorem is avoided. These vectors fields act as a 
second dark matter component in addition to the three scalar fields $\varphi _i$.
There have been other recent proposals for dark matter candidates with non-standard 
relationships between spin and mass dimension. In \cite{grumiller} \cite{grumiller1} a spin 1/2 dark matter 
candidate was proposed which has mass dimension 1. In the present case our vector fields,
$A^{\pm} _\mu$, $B^{\pm} _\mu$, have the same mass dimension -- 1 -- and statistics --
fermionic -- as the dark matter candidate in \cite{grumiller} \cite{grumiller1}, and only differ in the
value of spin -- 1 versus 1/2.  

\section{Summary and Conclusions}

We have given a model for phantom energy using a modification of the
graded Lie algebras models which attempted to give a more unified electroweak
theory, or Grand Unified theories. Despite interesting features of the original graded Lie 
algebra models ({\it e.g.} prediction of the Weinberg angle and having
both vectors and scalars coming from the same Lagrangian) they
had shortcomings. Chief among these was that if one used the correct SU(N/1) invariant
supertrace then some of the vector fields had the wrong sign for the kinetic energy term in 
the Lagrangian. In the original models the vector fields were associated with the even
generators of the algebra and the scalars fields were associated with the odd
generators. Here we took the reverse identification (scalar field $\rightarrow$ even generators
and vector field $\rightarrow$ odd generators) which led to the wrong sign kinetic energy
term coming from a scalar field rather than from a vector field. The wrong sign scalar field, $\varphi ^8$,
gives a model of phantom energy, while the other scalar fields, $\varphi ^i$, and the
vector fields, $A^a _\mu$, act as dark matter components. In the way our model is
formulated here all the fields are truly dark in that they have no coupling to any of
the Standard Model fields and would thus only be detectable via their gravitational interaction.
This would make the experimental detection of these dark fields impossible through
non-gravitational interactions. However the above is intended only as a toy model of how a phantom energy
field can emerge naturally from a gauge theory with a graded Lie algebra. A more experimentally testable
variation of the above toy model could have some coupling between 
the scalar and vector fields of the present model and the Standard Model fields. Such a
coupling could be introduced in a phenomenological fashion via some {\it ad hoc} coupling. A more interesting
option would be to consider some larger graded algebra, such as SU(N/1). Some of the fields could be given 
the standard assignment of even or odd generators ({\it i.e.} as in \eqref{graded-alg}) while others could be given
the assignment in \eqref{graded-alg2}. The fields given the standard assignment would give standard gauge 
fields, while fields given the non-standard assignment would give phantom energy and dark matter
fields. This would give a new type of ``Grand Unified Theory'' with the phantom energy and
dark matter fields replacing the extra gauge bosons of ordinary Grand Unified Theories.
Other authors \cite{wei} have used non-standard gauge groups such as SO(1,1) to give models of phantom energy.

An important feature of the above model is the assumption that the Grassmann
vector fields form permanently confined condensates. 
This was a crucial to our phantom energy model since it
leads to a condensate of the $A^\pm _\mu$ and $B^\pm _\mu$ fields. This in turn gave
a potential $V(\varphi ^8) =  \frac{v}{4}g^{2}(\varphi ^{8})^{2} $
for the $\varphi ^8$ field which was of the correct form to allow $\varphi ^8$ to act 
as phantom energy. Aside from the present application to phantom energy one
might try to use the above mechanism to generate standard symmetry break by starting with
a graded Lie algebra but using all vector fields rather than mixing vector and scalar. In 
this way some of the vector fields would be standard vector fields, while other would be
Grassmann vector fields. By the above mechanism the Grassmann vector fields would form
condensates which would then give masses to the standard vector fields {\it i.e.} one would have a
Higgs mechanism with only vector fields).  
 
One additional avenue for future investigation is to see if one could have a phantom energy model
with the original graded Lie algebra models ({\it i.e.} with vector fields assigned to
even generators and scalars to odd) but using the supertrace. One would then have the problem
of some of the vector fields having the wrong sign in the kinetic term, but this might
then give a phantom energy model with a vector rather than scalar field.

\begin{flushleft}
{\bf Acknowledgments} DS acknowledges 
the CSU Fresno College of Science and Mathematics for a sabbatical leave during 
the period when this work was completed.
\end{flushleft}


\end{document}